# Dynamic high pressure: Why it makes metallic fluid hydrogen

W. J. Nellis

Department of Physics

Harvard University

Metallic fluid H has been made by dynamic compression decades after Wigner and Huntington (WH) predicted its existence in 1935. The density obtained experimentally is within a few percent of the density predicted by WH. Metallic fluid H was achieved by multiple-shock compression of liquid $H_2$, which compression is quasi-isentropic and H is thermally equilibrated in experiments with 100 ns lifetimes. Quasi-isentropic means compressions are isentropic but with sufficient temperature and entropy to drive a crossover from liquid $H_2$ to degenerate fluid H at 9-fold compression and pressure P=140 GPa (1.4 Mbar). The metallic fluid is highly degenerate: $T/T_F \approx 0.014$, where temperature T=3000 K and Fermi temperature $T_F$ =220,000 K, respectively, at metallization density 0.64 mol $H/cm^3$.

Dynamic compression is achieved by supersonic, adiabatic, nonlinear hydrodynamics, the basic ideas of which were developed in the Nineteenth Century in European universities. Today dynamic compression is generally unfamiliar to the scientific community. The purposes of this paper are to present a brief review of (i) dynamic compression and its affects on materials, (ii) considerations that led to the sample holder designed specifically to make metallic fluid H, and (iii) an inter-comparison of dynamic and static compression with respect to making metallic H.

I. Introduction

In 1935 Wigner and Huntington (WH) predicted that solid insulating $H_2$ would undergo a transition to metallic H at a density 0.62 mol $H/cm^3$ (9-fold liquid-$H_2$ density), "very low temperatures", and a pressure greater than 25 GPa [1]. At sufficiently high densities, H is a metal because atomic $1s^1$ wave functions of adjacent atoms overlap to form an itinerant energy band. WH did not predict the pressure of their insulator-metal transition (IMT) because it was expected to be so large that its predicted value could not have motivated an experimental test with high-pressure technology of the 1930s. So WH simply calculated a lower bound on metallization pressure by assuming compressibility of hydrogen is constant with increasing pressure. They also realized there was no *a priori* reason why their IMT should occur. So, WH also predicted that, if their classic dissociative IMT does not occur, then at some high pressure solid $H_2$ would probably transform to a layer-like phase similar to graphite.

In 1956 when Stewart experimentally achieved 2 GPa in $H_2$ [2], making metallic hydrogen became a prime goal of the high-pressure community. In the late 1970s, 100 GPa (1 Mbar) pressures became accessible in a diamond-anvil cell (DAC). Static high-pressure research then began an extended period of growth.

In 1996 a metallic phase of H was achieved by dynamic compression at density $\rho$=0.64 mol $H/cm^3$, pressure P=140 GPa, temperature T $\approx$ 3000 K, and Fermi temperature $T_F$=220,000 K in experiments with ~100 ns lifetimes [3-5]. Because the melting temperature of hydrogen at 140 GPa is ~900 K [6], metallic H is metallic fluid H (MFH). In 2003 Fortov et al achieved similar results under dynamic compression [7]. MFH was made in both cases by multiple-shock compression of $H_2$, which is quasi-isentropic and thermally-equilibrated. MFH is degenerate condensed matter, i.e., $T/T_F \approx$ 0.014, typical of common metals at 300 K [8], which means MFH is at low temperatures with respect to the Fermi-Dirac distribution. However, because the temperature at metallization is finite, $\approx 3000$ K, the transformation from $H_2$ to H is observed as a crossover, rather than a sharp first-order transition predicted by WH.

Electrical conductivity of MFH is 2000/($\Omega$-cm), typical of strong-scattering, dense, monatomic, liquid metals H, N, O, Rb and Cs with electron scattering lengths comparable to average distance between adjacent atoms [9-11]. $H_2$ and H were probably the only hydrogen phases present in the dynamic compression experiments, as WH assumed in their model, because conductivity in the fluid at $\approx 3000$ K is not expected to permit stable $H_2$-$H_2$ intermolecular bonds to exist, as occurs in solid $H_2$ under static compression above ~240 GPa at ~300 K [12].

Theoretical calculations have shown warm, dense, fluid hydrogen to be quite complex. For example, transient dimers with lifetimes of ~$10^{-14}$ s have been observed in calculations for monatomic metallic fluid H [13]. This lifetime is comparable to a single vibrational period of a free $H_2$ molecule, which indicates the distinction between monomers and dimers is not clear-cut as densities and temperatures of MFH are approached [14].

Solid $H_2$ under static compression at the IMT density predicted by WH (0.62 mol H/cm$^3$ = 0.31 mol $H_2$/cm$^3$) is an insulator at pressure 73 GPa using the P-$\rho$ standard of Loubeyre et al [15] and is yet to metallize by dissociation at static pressures up to ~400 GPa, to date. The high-pressure metallic phase of hydrogen has been suggested to occur possibly by band overlap within a molecular solid [16] and via short-lived transient $H_2$ dimers in the fluid [14]. However, metallization of solid $H_2$ is yet to be observed and current experimental evidence indicates metallic fluid hydrogen is probably monatomic. Solid $H_2$ is not predicted to dissociate and probably metallize, until static pressure exceeds ~500 GPa [17], well beyond current DAC technology.

A major question from these results is why is it that metallic fluid hydrogen is made under dynamic compression at 140 GPa and ~3000 K, whereas at ~300 K solid metallic hydrogen is yet to be made at highest static pressures achieved in hydrogen to date, ~400 GPa. The likely answer is that (i) fast dynamic compression induces temperature T and entropy S, which drive dissociation from $H_2$ to H and (ii) the dynamic pressure pulse was shaped to produce quasi-isentropic compression of fluid H at sufficiently high pressures to make a degenerate metal. The dynamic-compression pulse weakens intra-molecular H-H bonds to dissociation in the warm fluid, in contrast to the formation of inter-molecular $H_2$-$H_2$ bonds in the cool solid, which strengthens solid molecular hydrogen and prevents dissociation to extremely high pressures. Lengthening the rise time of the dynamic pressure pulse from a few ps ($10^{-12}$ s) for a single-shock wave to 100 ns ($10^{-7}$ s) for a multiple-shock wave achieves lower temperatures and the higher densities needed to make degenerate MFH.

The basic ideas of dynamic compression were developed in the last half of the Nineteenth Century in European universities [18]. Today dynamic compression is generally unfamiliar to the scientific community, which impedes general understanding as to why dynamic compression makes MFH, whereas metallic solid hydrogen is yet to be made under static compression. The purposes of this paper are to present a brief review of (i) dynamic compression and its affects on materials, (ii) considerations that led to the sample holder designed specifically to make metallic fluid H, and (iii) an inter-comparison of dynamic and static compression with respect to prospects for making metallic H. An implication of this comparison is the possible existence in temperature-entropy space of a boundary region separating metal from nonmetal at 100 GPa pressures.

II. Basics of dynamic compression

Dynamic compression is achieved by supersonic, adiabatic, nonlinear hydrodynamics [18]. Metallic fluid H has been made because fast dynamic compression induces dissipation in the form of temperature T and entropy S. The extreme conditions needed to make MFH were generated by impact of a hypervelocity 20-g projectile with kinetic energy 0.5 MJ onto a cryogenic sample holder. Photographs of the experimental facility are published [19].

A shock wave is one type of dynamic-compression wave, a sharp step increase in state and hydrodynamic variables. The step increase is called a shock front, the region in which atoms/molecules are rapidly compressed with associated turbulent frictional heating and entropy production by dissociation of $H_2$. Material in the front equilibrates thermally via collisions. A shock wave travelling in a compressible Van der Waals fluid, such as Ar or $H_2$, compresses, heats, and disorders within the shock front (**Fig. 1**). Spatial and temporal widths of the front are the intervals in which shocked fluid comes into thermal equilibrium, ~nm and ps, respectively, in Van der Waals fluids. The rise time of a shock wave in liquid Ar is ~$10^{-12}$ s, orders of magnitude faster than rise time of pressure of ~s, which produces little dissipation in a DAC.

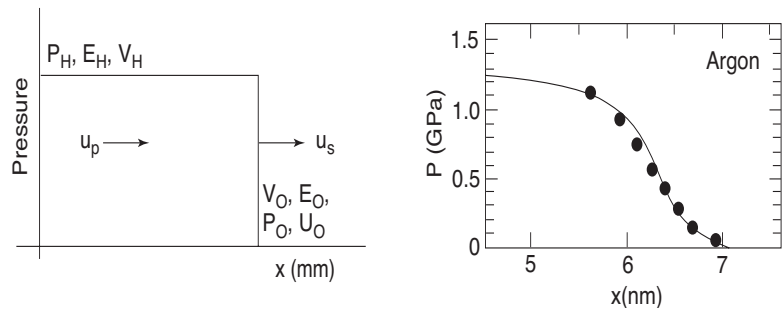

Fig 1. On left, schematic of single shock wave in liquid. $P_H$, $E_H$, and $V_H$ are state variables of shock pressure, internal energy on shock compression, and shock volume, respectively; zero-subscripted variables are initial values ahead of shock front; $u_s$ and $u_p$ are hydrodynamic flow variables of supersonic shock velocity and material velocity behind shock front, respectively. On right, calculated shock-front width in liquid argon showing rise in pressure to 1.2 GPa in ~nm, shock speed is 1.8 km/s, rise time from 0 to 1.2 GPa is ~$10^{-12}$ s. Full curve was calculated with atomistic molecular dynamics [20]; solid circles calculated with Navier-Stokes equations [21].

A dynamic-compression wave is a more general concept than a single shock. A multiple-shock wave, for example, is a sequence of step increases, each with fast rise-times of ~$10^{-12}$ s, separated by relatively long periods of $\approx 1$ ns to tens of ns of steady, virtually isentropic compression, as illustrated in **Fig. 2**. The total affect of such compression is quasi-isentropic. Thermodynamics achieved dynamically in liquid $H_2$ are tuned by tuning the hydrodynamics. Although the terms "shock" and "dynamic" do have different meanings, as described above, these terms are often used interchangeably. Both are used to prepare a material at extreme conditions for characterization by measured physical, chemical, and material properties.

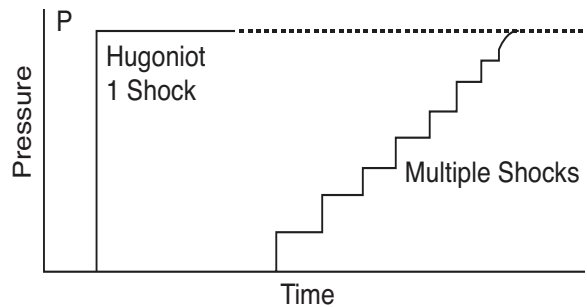

Fig. 2. Schematics of applied pressure histories of single-shock and multiple-shock compressions. Both compressions are adiabatic. However, multiple-shock compression is quasi-isentropic, as well. Dissipation T and S are induced in rapid step increases, while during times of nearly constant pressure compressions are essentially isentropic. By pulse shaping multiple-shock wave, T and S are reduced substantially from those caused by single sharp jump to same pressure.

The Rankine-Hugoniot (R-H) equations [22-24] conserve momentum, mass, and internal energy across the front of a single-shock wave and across each step increase in a multiple-shock wave (**Fig. 2**). They relate initial state and flow variables ahead of a travelling shock front, zero subscripts in Eqs. (1) to (3), to shock-state variables (H subscripts) and flow variables $u_s$ and $u_p$ behind that front. The conservation equations are:

$$P_H - P_0 = \rho_0 (u_s - u_0)(u_p - u_0) \qquad (1)$$

$$V_H = V_0[1-(u_p - u_0)/(u_s - u_0)] \qquad (2)$$

$$E_H - E_0 = 0.5(P_H + P_0)(V_0 - V_H) \qquad (3)$$

where mass density $\rho_H = 1/V_H$, $\rho_0 = 1/V_0$, $u_0$ is initial particle velocity in front of the shock wave, and "specific" means per gram g. The R-H equations describe an adiabatic process in which zero energy is transported into and out of the shock front during compression. Internal energy is deposited by shock compression itself.

The locus of states achieved by shock compression is called a Rankine-Hugoniot (R-H) curve, a Hugoniot, or a shock adiabat. The Hugoniot conservation equations apply whether or not shock-compressed material is equilibrated thermally. The question of thermal equilibrium must be decided independently of the R-H equations. Entropy in shock-compressed fluids is thermally equilibrated disorder in these discussions.

Shock impedance is defined as $Z = \rho_0 u_s$. From Eq. (1) with $u_0 = 0$ and $P_0 \ll P_H$, $P_H = Zu_p$. When a shock wave in one material crosses a boundary with a second material, P and $u_p$ adjust so that P and $u_p$ are continuous across the boundary and both materials move together with the same velocity $u_p$. That is, shock impedance $Z = P/u_p$ is matched across the boundary.

J. W. Strutt, Lord Rayleigh [25] and G. I. Taylor [26] developed complementary theories of instability growth along the interface between two materials with different densities under acceleration, which results in turbulent mixing along the boundary. Rayleigh-Taylor (R-T) theory is used to assess the likelihood that an interface between materials with different densities is hydrodynamically stable under large accelerations. The R-T instability is a major issue in Astrophysics and in Inertial Confinement Fusion (ICF).

III. Shock-induced T and S

WH predicted their IMT at 9-fold compression of liquid $H_2$. A compression this large cannot be achieved in states on the Rankine-Hugoniot curve, which for hydrogen has a maximum compression of only 4.3-fold over initial density [27,28]. However, compressions of $H_2$ substantially larger than 4.3-fold can be achieved by multiple-shock compression.

As discussed above, shock-induced S and T in the thin shock front dissociate $H_2$ to H. Multiple-shock compression of hydrogen in bulk is quasi-isentropic and can achieve relatively low temperatures and high densities needed to make degenerate metallic H (**Fig. 2**). In addition shock-induced dissipation in solids plays a role in the selection of an optimal solid anvil material with which to achieve multiple-shock compression. Shock-induced temperatures enable using conductivity measurements to observe the crossover from semiconductor to poor metal.

Hemholtz free energy is F=U-TS, where U is internal energy. F is minimized by maximizing entropy S. Shock-induced dissipation energy TS is energy that does not go into compression and in this sense can be considered "lost". If energy is "lost", that process is irreversible. The issue considered here is the split in dissipation between T and S that gives an optimal anvil material under dynamic compression, which depends primarily on the strength of interatomic interactions, which depends on anvil material.

By R-H equation (3) total energy deposited in shock compression is represented by the area of the triangle under the Rayleigh line, a line directly from $(P_0, V_0)$ to $(P_H, V_H)$. Reversible (isentropic) energy is represented by the area under the isentrope. Temperature increases with density along the isentrope are generally relatively small. Thus, isentropic pressure caused by temperature, called thermal pressure, is also generally small, in the sense that isentropic pressure of a fluid is nearly equal to isothermal pressure at 0-K. Irreversible thermal energy, or thermal dissipation energy, also goes into thermal pressure and is represented by the area between the Rayleigh line and the isentrope, the cross-hatched area in **Fig. 3**, which is plotted for aluminum [29].

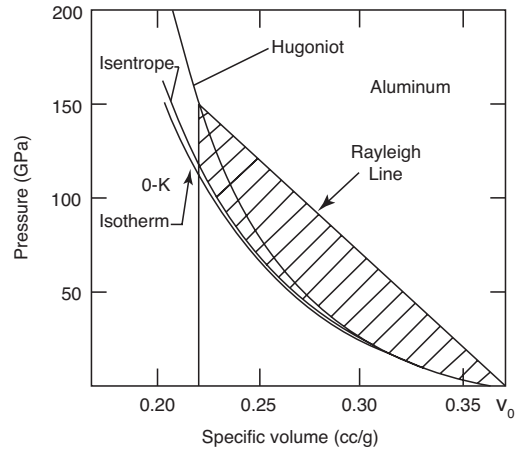

Fig. 3. Hugoniot, isentrope, and 0-K isotherm of Al [29], as well as Rayleigh line, plotted as pressure versus specific volume. $V_0$ is crystal volume at ambient. Isentrope and isotherm are reversible; Hugoniot is not. Irreversible thermal-dissipation energy of Hugoniot is represented by cross-hatched area.

III.1 Simple metals: Al

Hugoniot pressure $P_H$, isentropic pressure $P_s$ and 0-K isothermal pressure $P_0$ of Al are plotted up to 150 GPa versus specific volume [29] in **Fig. 3**. In this case heating by isentropic compression is relatively small and shock-induced disorder (S) of the Al lattice has a relatively weak affect on pressure. Irreversible thermal pressure is $P_{th} = P_H - P_s$ as a function of volume. In short, simple metals, such as Al, are relatively compressible and have modest shock heating.

III.2 Rare-gas liquids: Ar

Atoms in rare-gas liquids interact via effective Van der Waals pair potentials, which are a relatively weak. Thus, fluid Ar, for example, compresses rapidly with a substantial volume collapse, and equilibrates thermally to a relatively high T. As seen in **Fig. 4** [30], only ~10% of total shock energy is reversible; the rest is irreversible thermal heating, which causes substantial thermal pressure. For a liquid sample and fluid shock state shock-induced entropy is relatively small. The "softening" on the Ar Hugoniot at ~50 GPa is caused by shock energy absorbed by thermal excitation of electrons across the mobility gap between valence and conduction bands. Ar is a fluid, which means pressure is isotropic.

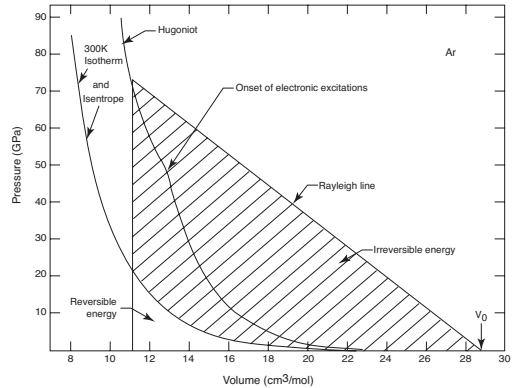

Fig. 4. Hugoniot, isentrope, and 0-K isotherm of liquid Ar plotted as pressure versus molar volume [30].

Ar and Al are examples of materials that produce significant irreversible shock heating. Optimal anvil material would produce substantial shock-induced dissipation in the form of disorder, with relatively little shock heating. In this case the Hugoniot would have a smaller component of thermal pressure and pressure release curves would be relatively close to the Hugoniot. Such a solid would be optimal anvil material in the sense that it would be more nearly reversible on decompression and re-shock, as happens to anvils under shock reverberation.

III.3 Strong oxides: Sapphire ($Al_2O_3$)

Sapphire (single-crystal $Al_2O_3$) was used as the anvil material between which to reverberate a shock wave in liquid $H_2$. In this section experiments needed to qualify sapphire, or any solid, for use as an anvil material under dynamic compression are discussed, including dynamic strength, equation of state, and electrical conductivities. In addition, results of static compression experiments on possible anvil materials at comparable pressures and temperatures can provide important supplemental information for dynamic experiments.

Shock-induced dissipation in sapphire primarily damages the sapphire lattice with its strong ~5-eV Al-O bonds plus some shock-induced heating. It is unlikely that shock-induced damage in solid sapphire equilibrates thermally below ~300 GPa because of strong bonds and experimental lifetimes of ~100 ns or less. Shock dissipation is heterogeneous at 45 GPa [31] and probably remains so to pressures well above 100 GPa. Pressures relavent to hydrogen are 90 to 180 GPa. Above ~300 GPa shock-induced dissipation in sapphire is primarily thermal.

Sapphire is a strong solid that supports an elastic shock wave up to the Hugoniot elastic limit (HEL), the shock stress at which a strong material fails plastically in the direction of shock propagation. Because a solid has strength, longitudinal and transverse components of stress are unequal and stress is anisotropic. Just above the HEL the elastic wave is faster than the plastic wave and thus a two-wave shock structure develops. As shock stress increases above the HEL, the elastic-wave velocity remains essentially constant and the plastic-wave velocity increases with increasing shock stress. At some high shock stress $P_{OD}$, velocity of the plastic shock wave $u_{sOD}$ becomes equal to that of the elastic shock $u_{HEL}$. Above $P_{OD}$ the elastic shock is overdriven by the plastic wave and only a single shock wave is observed in shock-wave profiles. A material can have elastic strength above $P_{OD}$, but elastic strength is not evident in shock-wave structure above $P_{OD}$. Because a two-wave shock structure in strong sapphire might need to be taken into account in analysis of hydrogen electrical-conductivity data, the stress regime of the two-wave structure was measured.

Shock-wave profiles of hexagonal sapphire crystals cut with seven different crystal orientations in the direction of shock propagation, including c-cut used in the hydrogen experiments [4], were measured at 16, 23, and 86 GPa [32]. The HEL and $P_{OD}$ of all seven orientations are about ~18 GPa and ~90 GPa, respectively. Because the two-wave structure is overdriven at 90 GPa [32] and hydrogen conductivities were measured at 93 GPa and higher [4], affects of a two-wave shock structure are negligible at the relevant pressures and thus two-wave affects in sapphire were neglected in analysis of the hydrogen conductivity experiments.

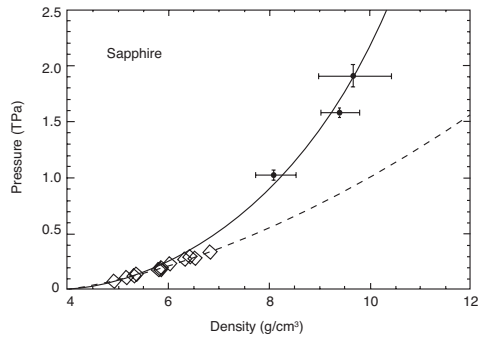

Fig. 5. Hugoniot of sapphire (open diamonds: [33,34]; points with error bars: [35]; dashed curve: extrapolation of combined fit to $u_s$-$u_p$ data in [33,34]; calculated 0-K isotherm [36] (not shown) is at somewhat lower pressures than dashed curve.

The measured sapphire Hugoniot is plotted up to 1.9 TPa (19 Mbar) [33-35] in **Fig. 5**. The pressure-density curve of sapphire was calculated up to ~800 GPa at 300 K [36]. The dashed curve is an extrapolation of the combined linear fit to $u_s$-$u_p$ Hugoniot data below 0.340 TPa [33,34]. The calculated 300-K isotherm is at somewhat lower pressures than the measured Hugoniot below ~300 GPa because of a modest component of thermal shock pressure. The fact that measured and calculated Hugoniots of sapphire are quite close to one another below ~300 GPa implies that crystal damage/disorder is the dominant shock dissipation mechanism. At a shock pressure of 150 GPa measured electrical conductivity of shocked sapphire is a factor of $10^4$ smaller than that of MFH [37], which means electrical conduction between two metal electrodes inserted through an $Al_2O_3$ anvil to touch MFH (see below) was negligible compared to conduction through MFH.

Under static compression in a DAC at 300 K only disordered $Al_2O_3$ is produced at static pressures in the range 120 to 180 GPa and static laser-heating at ~2000 K is needed to observe phase transitions in heavily disordered $Al_2O_3$ [38]. Those static pressures and temperatures are essentially the same as those of the hydrogen electrical conductivity experiments under dynamic compression [4], with one major difference. Dynamic temperatures and pressures had a duration of ~100 ns, whereas static laser-heating at high pressures probably had a lifetime of several minutes. The implication for dynamic compression of those DAC experiments is that substantial shock-dissipation energy is readily absorbed by disordering the $Al_2O_3$ lattice above ~100 GPa, which creates substantial damage/disorder at the expense of heating. On release of shock pressure from a point on the Hugoniot the fact that the Hugoniot, 0-K isotherm, and isentrope in compression are nearly coincident up to ~200 GPa for $Al_2O_3$ [39] suggests that thermal pressures are relatively small and thus pressure-release isentropes from pressures up to 200 GPa on the Hugoniot are nearly coincident with their Hugoniot.

IV. Isentropic compression

Isentropic energy $E_S$ achieved by slow, reversible, isentropic compression of material from volume $V_1$ to $V_2$ is

$$E_S = -\int_{V_1}^{V_2} P_S(V,T)dV \qquad (4)$$

Isentropic compression produces less heating and more compression for a given pressure change than does fast adiabatic shock compression. Temperature and density at a given pressure can be tuned by tuning the relative amounts of shock and isentropic compression caused by an applied pressure pulse. Multiple-shock compression is one way to tune the relative amounts of both. In a multiple-shock wave, the first wave is generally a shock and the remainder of the shocks are essentially isentropic. Thus, a multiple shock is often called quasi-isentropic.

V. Shock reverberation

Multiple-shock compression of liquid $H_2$ is readily obtained by reverberating a shock wave in compressible liquid $H_2$ contained between relatively incompressible high-density solid anvils. Multiple shocks mean multiple step increases in pressure (**Fig. 2**). Multiple-shock compression is quasi-isentropic; that is, mostly isentropic but with enough shock-induced dissipation T and S to dissociate $H_2$ to H, while producing sufficiently high density to make metal H at sufficiently low temperature that the metal is degenerate, $T/T_F \ll 1$. Because WH predicted an IMT from $H_2$ to H requires 9-fold compressed liquid $H_2$, many step compressions are needed to achieve such a large total compression. It is advantageous to use anvil materials that have small thermal pressures under shock compression because Hugoniot and pressure-release isentropes of such materials are nearly identical and reversible. Further, it is desirable that the anvil be an electrical insulator during shock reverberation so that electrical conductivities of hydrogen can be measured by inserting metal electrodes through an anvil to contact the hydrogen sample without current being conducted between anvils through the shocked anvil.

Sapphire is an ideal anvil material for several reasons. The sapphire Hugoniot is an excellent equation of state (EOS) [33,34] with which to simulate shock reverberation. The density of strong sapphire is 4.00 g/cm$^3$, 56 times greater than the density of liquid $H_2$, 0.071 g/cm$^3$, which means many shock reverberations are needed to bring liquid $H_2$ initially at atmospheric pressure quasi-isentropically up to MFH at 140 GPa. Measured electrical conductivity of sapphire at 140 GPa is 5 orders of magnitude smaller than 2000/(Ω-cm) of MFH [37], which means sapphire is effectively an electrical insulator compared to MFH, as is needed.

Figure 6 is a schematic of the cryogenic sample holder used to make MFH. The sample chamber is filled with liquid $H_2$ at 20 K under $H_2$-gas pressure of ~900 torr to suppress boiling of liquid $H_2$ on the saturation curve. In particular, the volume between two sapphire discs was filled with liquid $H_2$. A four-probe electrode configuration was used to insert electrical current into the liquid $H_2$ sample and to measure the voltage drop across the $H_2$ sample once it became conducting on shock reverberation. The cryogenic system to fill and maintain liquid $H_2$ (not shown) was controlled remotely because no one is allowed in the room with the two-stage light-gas gun when the gun is fired. Voltages were measured on fast digitizers with ns resolution. The 20-g impactor was an Al or Cu plate contained in a Lexan sabot (not shown) launched by a 20-m long two-stage light-gas gun to velocities in the range 5.6 to 7.3 km/s. Impactor velocity $u_I$ was measured inflight by two 50-ns X-ray pulses separated spatially by 30 cm and temporally by ~40 μsec at maximum $u_I$. Impactor velocity $u_I$ was chosen to give a desired pressure in hydrogen after the wave reverberations are complete. The geometrical factor that relates measured electrical resistance to electrical resistivity was calculated with a Monte Carlo code.

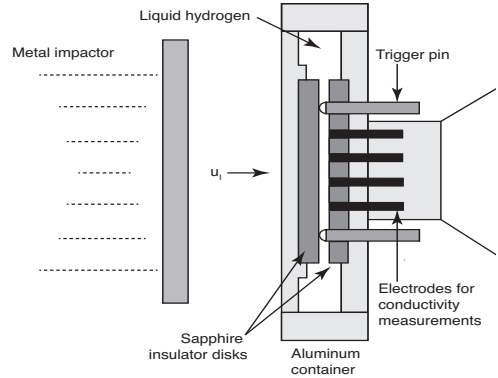

Fig. 6. Dynamic compression experiment on liquid $H_2$ initially at 20 K. Liquid $H_2$ diameter and thickness were 28 mm and 0.5 mm, respectively.

Shock hydrodynamics of wave reverberation in the ($Al_2O_3$-liquid $H_2$-$Al_2O_3$) "sandwich" in the center of the sample holder in Fig. 6 are illustrated in P-$u_p$ space in Fig. 7(a). P and $u_p$ are the only variables in the R-H equations (1)-(3) that are continuous across an interface between two materials. Fig. 7(b) is a plot of pressure history at the midpoint of the hydrogen layer. Both 7(a) and 7(b) were calculated with a one-dimensional hydrodynamics computational code. The numbers along the two P($u_p$) curves in Fig. 7(a) and on the steps in Fig. 7(b) indicate the first shock, second shock, etc of the reverberation. Fig. 7(b) also indicates the time duration of each pressure jump during the compression process. Shock velocity increases with shock pressure and so the duration of each step decreases as shock pressure increases and sample thickness decreases during the compression process.

In the experiment at 140 GPa, illustrated in Fig. 7(a), the impact of the Cu plate with the Al container generates a 100-GPa planar shock wave in Al, which after transit of the Al layer shocks up to 140 GPa on reflection off the first sapphire disk. When the 140-GPa shock wave in $Al_2O_3$ reaches the interface with liquid $H_2$, the 140 GPa shock pressure releases down the $Al_2O_3$ release curve on the right side of Fig. 7(a) until it intersects the shock Hugoniot of liquid $H_2$ at 5 GPa (point 1). At point 1 the first shock in liquid $H_2$ reflects up in amplitude off the sapphire disk on the right side of the sample chamber in Fig. 6 and then travels back toward the left from whence it came. The curve from points 1 to 2 is called the reflected or reshock Hugoniot of hydrogen. At point 2 the wave in hydrogen coming from point 1 reflects off the sapphire disc on the left of Fig. 6(a) and heads back toward point 3 on the right side of Fig. 6(a). The shock wave then reverberates back and forth in hydrogen between the two anvils until the pressure in fluid hydrogen reaches 140 GPa, which holds constant for another 50 ns. Since the speed of voltage signals in the 0.5 mm diameter cryogenic coaxial cables is essentially the speed of light, the voltage signals are captured on the digitizers prior to the explosion of the holder. A new sample holder was built for each experiment.

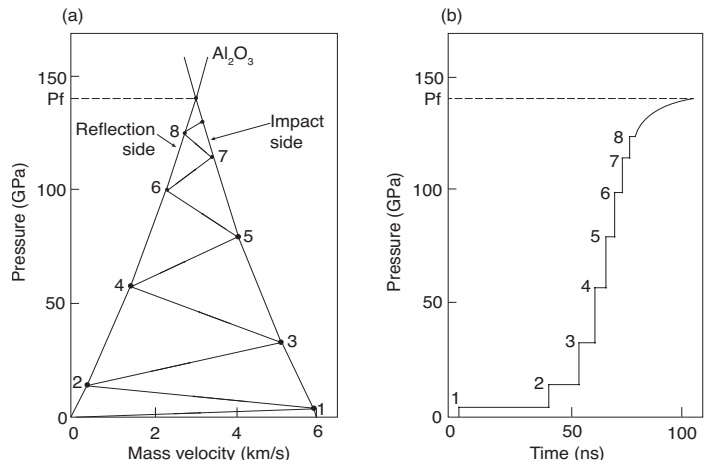

Fig. 7. (a) Hugoniot and pressure-release isentrope of $Al_2O_3$ in P-$u_p$ space starting from 140 GPa, achieved by a Cu impactor striking cryogenic target in Fig. 6 at velocity 5.6 km/s. These curves were calculated with one-dimensional hydrodynamics simulation code. (b) Pressure versus time at midpoint of hydrogen layer initially 0.5 mm thick. Eight distinct shocks are resolved in Fig. 7(b) before the transit times of the shocks become so small that they "smear" into a continuum. New sample holder was built for each sample pressure $P_f$.

In this way the shock pressure in hydrogen reverberates between the two sapphire discs until the pressure in hydrogen reaches final pressure $P_f$, which also equals the initial shock pressure $P_i$ in $Al_2O_3$. $P_i$ is determined simply by matching shock impedances of Cu impactor, Al wall, and $Al_2O_3$, independent of the hydrogen EOS. The Hugoniots of the Cu impactor, Al wall, and $Al_2O_3$ anvil were all measured previously and are EOS standards [34,40].

Liquid $H_2$ is compressed, heated and dissociated in the narrow width of the shock front (Fig. 1) as the various step increases in pressure traverse the sample (Fig. 2). To make bulk metallic H, overlap of electron wave functions on adjacent atoms must be compressed sufficiently to form an itinerant energy band. To do that, the pressure pulse on the sample in bulk must be applied sufficiently slowly to compress the atoms prepared in the front into a dense degenerate metal. To do that liquid-$H_2$ was multiply-shocked by shock reverberation, as illustrated by each step jump in pressure in Fig. 7(b). Shocks after the first one are effectively isentropes, which achieve lower temperatures and higher densities than shocks.

Because of finite T produced by dynamic compression, electrons are thermally excited from the valence band to the conduction band of semiconducting fluid hydrogen, which enables measurements of electrical conductivities. Figure 8 is a plot of resistivity versus pressure in the range 93 to 180 GPa. Each of those pressures is the final pressure $P_f$ achieved on completion of a reverberation process illustrated in Fig. 7, for each of ten experiments. From 93 to 140 GPa electrical conductivity is thermally activated, and so electrical resistivity decreases logarithmically. In this range the energy gap is closing with pressure-induced compression and temperature is increasing with increasing pressure. The net affect is to increase the number of electron carriers to that of a metal. From 140 to 180 GPa the mobility gap is closed and electrical conductivity is constant at $2000/(\Omega\text{-cm})$, typical of a strong scattering liquid metal.

Hydrogen and deuterium points fall on a common curve in Fig. 8 because temperatures of the data points are in the range 1500 to 3000 K. Quantum effects of the protons and deuterons are expected to be destroyed by these high temperatures. However, a temperature of 3000 K with a free-electron Fermi temperature of 220,000 K means the electron gas is quantum in nature with a free-electron degeneracy factor of $T/T_F \approx 0.014$.

From measured conductivities in the semiconductor range between 90 and 140 GPa, the density dependence of the mobility gap, $E_g(\rho)$ was derived. Metallization at finite T in MFH occurs at the density $\rho_{met}$ at which $E_g(\rho_{met}) \approx k_B T$, where $k_B$ is Boltzmann's constant and $\rho_{met}$=0.32 mol $H_2/cm^3$=0.64 mol $H/cm^3$. $\rho_{met}$ is weakly sensitive to whether T is the temperature of MFH (~0.3 eV) or 0 K [4].

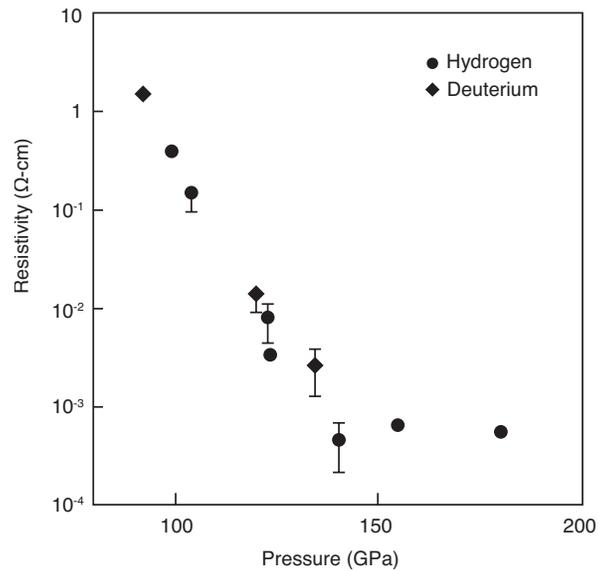

Fig.8. Electrical resistivity of fluid H versus multiple-shock pressure [4]. Corresponding temperatures are 1500 to 3000 K. Density at end of crossover from fluid semiconductor to fluid metal is 0.32 mol $H_2/cm^3$=0.64 mol $H/cm^3$. Free-electron Fermi temperature $T_F$ at this density is $T_F$=220,000 K. Metallic fluid H is degenerate ($T/T_F \approx 0.014$). Wigner and Huntington predicted their IMT from insulating $H_2$ to metallic H at 0.62 mol $H/cm^3$ [1].

VI. Discussion

Shock-induced temperature and entropy have been shown to have a major affect on whether or not electrons are delocalized in highly condensed fluid hydrogen. T and S generated by supersonic hydrodynamics drive metallization under certain conditions. The Hugoniot curve (single-shock compression), multiple-shock compression curve, and the 0-K isotherm are illustrated schematically in Fig. 9. It is not possible to metallize hydrogen under dynamic compression on its Hugoniot because sufficient T is generated on increasing shock pressure that maximum shock compression of $H_2$ on its Hugoniot is only 4.3 fold [27,28]. According to WH the $H_2$-to-H IMT occurs at a compression of 7.4-fold or 9.0-fold of solid-$H_2$ or liquid-$H_2$ density, respectively, which is too large to be achieved on the Hugoniot. At ~300 K metallic hydrogen has yet to be made at pressures up to ~400 GPa. Under multiple-shock compression MFH has been made at 9.0-fold compression of liquid-$H_2$ density, ~3000 K, and 140 GPa. These results suggest that (i) too much T and S are generated under single-shock compression of liquid $H_2$ to form MFH in condensed matter on the Hugoniot, (ii) insufficient S has been generated to date at ~300 K up to ~400 GPa to form metallic solid H near the 300-K isotherm, and (iii) sufficient T and S have been generated under multiple-shock compression to make MFH at 0.64 mol H/cm$^3$, essentially the density predicted by WH, 0.62 mol H/cm$^3$. These observations constitute a sparse data set that suggests the existence of a boundary region in T-S space of hydrogen between metal and nonmetal at 100 GPa pressures.

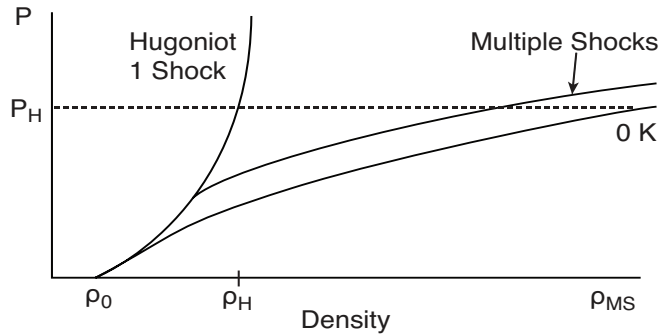

Fig. 9. Schematic in pressure-density space of single-shock (Hugoniot), multiple-shock compression as in Fig. 6, and the 0-K isotherm as in a DAC.

Thermodynamic states in hydrogen achieved by dynamic compression can be tuned by choice of anvil material, wall material, impactor material and impact velocity (Fig. 6). For example, higher pressures and lower temperatures can be achieved in hydrogen by using anvil materials with higher mass density than sapphire. $Gd_3Ga_5O_{12}$ (GGG), for example, has density 7.1 $g/cm^3$, which is large compared to 4.0 $g/cm^3$ of sapphire. Additional diagnostic probes, other than electrical conductivity, need to be developed to explore this regime. For example, sapphire becomes opaque at a shock pressure of 130 GPa. Thus, dense transparent anvil/windows and optical spectrometers need to be obtained to measure temperature of dynamically compressed low-density liquids and solids by thermal radiation spectra and to perform laser-spectroscopy experiments on those shock-compressed materials.